\documentclass{article}

\usepackage{amsmath, amsfonts, amssymb}
\usepackage{booktabs}
\usepackage{enumerate}
\usepackage{graphicx}
\usepackage{hyperref}
\usepackage{multirow}
\usepackage[square, numbers, sort&compress]{natbib}

\newcommand{\dif}{\ensuremath{\,\mathrm{d}}}
\DeclareMathOperator{\E}{\mathbb{E}}
\newcommand{\indicator}{\ensuremath{\mathbb{I}}}
\newcommand{\bigcdot}{\boldsymbol{\cdot}}

\begin{document}

\title{{\Large Pairwise accelerated failure time regression models for infectious disease transmission in close-contact groups with external sources of infection}}
\author{Yushuf Sharker,$^1$ Zaynab Diallo,$^2$ Wasiur R. KhudaBukhsh,$^3$ \\ and Eben Kenah$^{2,*}$}
\date{{\small
  $^1$ Data Sciences Institute, Takeda Pharmaceuticals USA \\
  $^2$ Biostatistics Division, College of Public Health, The Ohio State University \\
  $^3$ School of Mathematical Sciences, University of Nottingham \\
  $^*$ Correspondence: kenah.1@osu.edu
}}
\maketitle








\begin{abstract}
  Many important questions in infectious disease epidemiology involve the effects of covariates (e.g., age or vaccination status) on infectiousness and susceptibility, which can be measured in studies of transmission in households or other close-contact groups.
  Because the transmission of disease produces dependent outcomes, these questions are difficult or impossible to address using standard regression models from biostatistics.
  Pairwise survival analysis handles dependent outcomes by calculating likelihoods in terms of contact interval distributions in ordered pairs of individuals.
  The contact interval in the ordered pair $ij$ is the time from the onset of infectiousness in $i$ to infectious contact from $i$ to $j$, where an infectious contact is sufficient to infect $j$ if they are susceptible.
  Here, we introduce a pairwise accelerated failure time regression model for infectious disease transmission that allows the rate parameter of the contact interval distribution to depend on infectiousness covariates for $i$, susceptibility covariates for $j$, and pairwise covariates.
  This model can simultaneously handle internal infections (caused by transmission between individuals under observation) and external infections (caused by environmental or community sources of infection).
  In a simulation study, we show that these models produce valid point and interval estimates of parameters governing the contact interval distributions.
  We also explore the role of epidemiologic study design and the consequences of model misspecification.
  We use this regression model to analyze household data from Los Angeles County during the 2009 influenza A (H1N1) pandemic, where we find that the ability to account for external sources of infection is critical to estimating the effect of antiviral prophylaxis. \\[5pt]
  \textbf{Keywords:} accelerated failure time model, infectious disease epidemiology, secondary attack risk, survival analysis
\end{abstract}


\section{Introduction}
Many important questions in infectious disease epidemiology involve the effects of covariates on the risk of transmission. 
Studies of households and other groups of close contacts are one of the most valuable sources of information about mechanisms and risk factors for transmission~\citep{frost1938familial, fox1974family}.
They have been used to understand many recent emerging and re-emerging epidemics, including 
pertussis~\citep{halloran2003estimating}, SARS coronavirus~\citep{goh2004secondary}, 2009 pandemic influenza A (H1N1)~\citep{cauchemez2009household, morgan2010household, france2010household, carcione2011secondary, savage2011assessing}, MERS coronavirus~\citep{drosten2014transmission}, Ebola virus disease~\citep{glynn2018variability, reichler2018household}, norovirus~\citep{marsh2018unwelcome, tsang2018transmissibility}, hand-foot-and-mouth disease~\citep{hoang2019transmission}, cryptosporidium~\citep{korpe2019case}, measles~\citep{banerjee2020containing}, and COVID-19~\citep{bi2020epidemiology, li2020characteristics}.
The vast majority of these studies are analyzed using logistic regression or other standard models for binary outcomes.
These implicitly attribute all subsequent infections in the household to the primary case, failing to account for multiple generations of transmission within the household~\citep{sharker2021estimating} and for the ongoing risk of infection from sources outside the household.

The transmission of disease creates dependent outcomes in different individuals because the infection of one individual alters the risk of infection in other individuals.
These dependencies are not accounted for in standard regression models, which assume independent or conditionally independent outcomes in different individuals~\citep{HalloranStruchiner1991}.
Attempts to account for these dependencies using robust variance are ineffective because they do not address the bias in the underlying point estimates~\citep{sharker2021estimating}.
When exposure to infection differs systematically in susceptible individuals with different covariate values, estimates of covariate effects on susceptibility that do not account for this dependence can be biased even in randomized trials~\citep{morozova2018risk, eck2022randomization}.

In pairwise survival analysis, dependent outcomes in individuals are handled by analyzing failure times in ordered pairs of individuals~\citep{kenah2008generation, kenah2011contact}. 
In the ordered pair $ij$, the \emph{contact interval} is the time from the onset of infectiousness in $i$ to infectious contact from $i$ to $j$, where an infectious contact is defined to be a contact sufficient to infect $j$ if they are susceptible. 
The survival function of the contact interval distribution can be used to calculate the probability of transmission from $i$ to $j$ over any given time interval during which $i$ is infectious.
The probability of transmission to $j$ during the entire infectious period of $i$ is called the \emph{secondary attack risk} (SAR).
The hazard function of the contact interval distribution is a measure of the relative infectiousness of $i$ as a function of time elapsed since the onset of infectiousness, which is often called the \emph{infectiousness profile}.
The contact interval from $i$ to $j$ is right-censored if $i$ recovers from infectiousness prior to infectious contact with $j$, if $j$ is infected from a source other than $i$, or if observation of the pair $ij$ ends before $j$ is infected. 
When who-infected-whom is observed, standard parametric and nonparametric methods from survival analysis can be used to estimate the contact interval distribution. 
When who-infected-whom is not observed, parametric likelihoods can be integrated over all possible transmission trees~\citep{kenah2011contact} or nonparametric estimates from all possible transmission trees can be averaged using an expectation-maximization algorithm~\citep{kenah2013non}. 

For transmission from an infectious individual $i$ to a susceptible individual $j$, there are three possible types of covariates: Covariates for $i$ could affect their infectiousness, covariates for $j$ could affect their susceptibility, and pairwise covariates (e.g., membership in the same household) could affect the risk of transmission independently of the infectiousness of $i$ or the susceptibility of $j$. 
Estimation of these effects can be used to design and evaluate public health responses to epidemics~\citep{HalloranStruchinerLongini1997, HalloranLonginiStruchiner}.

To allow estimation of covariate effects on the hazard of transmission, Kenah~\citep{kenah2015semiparametric} developed a semiparametric regression model in which the hazard of infectious contact from $i$ to $j$ was 
\begin{equation}
    h_{ij}(\tau) = e^{\beta^\top X_{ij}} h_0(\tau),
\end{equation}
where $\beta$ is a coefficient vector, $h_0(\tau)$ is an unspecified baseline hazard for the contact interval, and $X_{ij}$ is a vector that can include infectiousness covariates for $i$, susceptibility covariates for $j$, and pairwise covariates. 
This allows point and interval estimation of hazard ratios associated with covariates, but it still assumes that all transmissions occur between individuals under observation. 
The inability to account for external sources of infection is a limitation that must be addressed before pairwise survival analysis can become a practical tool for infectious disease epidemiology.

In this paper, we develop a pairwise accelerated failure time (AFT) regression model that allows the rate parameter of the contact interval distribution to depend on covariates while accounting simultaneously for the risk of infection from internal and external sources.
We use counting process theory and a simulation study to show that parameter estimates from this model are consistent and asymptotically normal.
Our simulation study also highlights the critical role of epidemiologic study design in parameter estimation.
We apply the pairwise AFT regression model to household surveillance data collected by the Los Angeles County Department of Public Health during the 2009 influenza A (H1N1) pandemic, and we find that accounting for external infection improves statistical power to estimate the effect of antiviral prophylaxis. 
The pairwise AFT model has the potential to become an important new statistical tool in infectious disease epidemiology, with potential applications including the design and analysis of vaccine trials, outbreak investigations, and the analysis of contact tracing data or household studies. 

\subsection{Stochastic S(E)IR models}
The pairwise AFT regression model is based on a very general stochastic model of transmission.
At any time, each individual $i \in \{1, \ldots, n\}$ is in one of four states: susceptible (S), exposed (E), infectious (I), or removed (R).  
Person $i$ moves from S to E at their \emph{infection time} $t_i$, with $t_i = \infty$ if $i$ is never infected.  
After infection, $i$ has a \emph{latent period} of length $\varepsilon_i$ during which they are infected but not infectious.  
At time $t_i +\varepsilon_i$, $i$ moves from E to I, beginning an \textit{infectious period} of length $\iota_i$. 
At time $t_i+ \varepsilon_i + \iota_i$, $i$ moves from I to R, where they can no longer infect others or be infected. 
The latent period $\varepsilon_i$ is a nonnegative random variable, and the infectious period $\iota_i$ is a strictly positive random variable.
We assume that the latent and infectious periods both have finite mean and variance.
The time elapsed since the onset of infectiousness in $i$ at time $t_i + \varepsilon_i$ is called the \emph{infectious age} of $i$. 
An SIR model is an SEIR model with latent period $\varepsilon_i = 0$ for all $i$. 

After becoming infectious at time $t_i + \varepsilon_i$, person $i$ makes infectious contact with $j \neq i$ at time $t_{ij} = t_i + \varepsilon_i + \tau^*_{ij}$. 
The \textit{infectious contact interval} $\tau^*_{ij}$ is a strictly positive random variable with $\tau^*_{ij} = \infty$ if infectious contact never occurs.  
Because infectious contact can only occur while $i$ is infectious, we always have $\tau^*_{ij} \in (0, \iota_i]$ or $\tau^*_{ij} = \infty$. 
Because we define infectious contact to be sufficient to infect a susceptible person, $t_j \leq t_{ij}$ for all $i$ and $j$. 

An \emph{internal infection} occurs when an individual is infected by another individual in the observed population. 
For each ordered pair $ij$, let $C_{ij} = 1$ if infectious contact from $i$ to $j$ is possible and $C_{ij} = 0$ otherwise. 
For example, $C_{ij}$ could be an entry in the adjacency matrix for a contact network through which infection is transmitted.
We do not assume that $C_{ij} = C_{ji}$, so this network could contain directed edges.
The infectious contact interval $\tau^*_{ij}$ is generated as follows: 
A \textit{contact interval} $\tau_{ij}$ is drawn from a failure time distribution with hazard function $h_{ij}(\tau)$. 
If $\tau_{ij}\leq\iota_i$ and $C_{ij} = 1$, then $\tau^*_{ij} = \tau_{ij}$.  
Otherwise, $\tau^*_{ij} = \infty$. 
The hazard function $h_{ij}(\tau)$ is the instantaneous infectiousness of $i$ at time $\tau$ after the onset of infectiousness.
The cumulative hazard function is $H_{ij}(\tau) = \int_0^\tau h_{ij}(u) \dif u$, and the secondary attack risk is $1 - S_{ij}(\iota_i)$ where $S_{ij}(\tau) = \exp(-H_{ij}(\tau))$ is the survival function.

An \emph{external infection} occurs when an individual is infected from a source outside the observed population, such as an environmental source or a community source (i.e., an individual who is not under observation). 
Let $C_{0j}$ indicate whether individual $j$ is at risk of external infectious contact. 
Let the \emph{external infectious contact time} $t^*_{0j}$ denote the first time that an individual $j$ receives infectious contact from outside the observed population, with $t^*_{0j} = \infty$ if this never occurs. 
We assume that the external infectious contact time is generated as follows: 
An \emph{external contact interval} $t_{0j}$ is drawn from a failure time distribution with hazard function $h_{0j}(t)$. 
If $C_{0j} = 1$, then $t^*_{0j} = t_{0j}$.
Otherwise, $t^*_{0j} = \infty$.
For simplicity, we assume the external source is always infectious.
This assumption could be relaxed by defining an infectiousness onset time $t_0 + \varepsilon_0$ and infectious period $\iota_0$ for the external source.

\subsection{Exposure and infectious sets}
For each internal infection $j$, let $v_j$ denote the index of his or her infector.
Let $v_j = 0$ if $j$ is an external infection and $v_j = \infty$ if $j$ is not infected. 
When $v_j$ is observed for all infected $j$, we say that \emph{who-infected-whom} is observed. 
Otherwise, we say that who-infected-whom is not observed even if $v_j$ is observed for a subset of infected $j$. 

The \emph{exposure set} of an individual $j$ is 
\begin{equation}
  \mathcal{W}_j = \{i < \infty: (t_i + \varepsilon_i < t_j \text{ or } i = 0) \text{ and } C_{ij} = 1\},
\end{equation}
which is the set of all sources of infection to whom $j$ was exposed while susceptible.
The \emph{infectious set} of individual $j$ is the set of all possible $v_j$, which we denote $\mathcal{V}_j$.
Given $C_{ij}$ and transfer times between the states (S, E, I, and R), we must have
\begin{equation}
  \mathcal{V}_j \subseteq \{i < \infty: (t_i + \varepsilon_i < t_j \leq t_i + 
  \varepsilon_i + \iota_i \text{ or } i = 0) \text{ and } C_{ij} = 1\}.
  \label{eq:Vj}
\end{equation}
If the infector of $j$ is known (perhaps through pathogen genome sequences), then $\mathcal{V}_j = \{v_j\}$. 
If $j$ was not infected, then $\mathcal{V}_j = \varnothing$ (the empty set).

\subsection{Infectious disease data}
Our epidemiologic data contain the times of all $\text{S} \rightarrow \text{E}$ (infection), $\text{E} \rightarrow \text{I}$ (infectiousness onset), and $\text{I} \rightarrow \text{R}$ (removal) transitions in the observed population between time $0$ and a time $T$ that is a stopping time with respect to the observed data. 
For all ordered pairs $ij$ in which $i$ is infected or $i = 0$, we observe $C_{ij}$. 

The contact interval $\tau_{ij}$ can be observed only if $j$ is infected by $i$ at time $t_{ij} = t_i + \varepsilon_i + \tau_{ij}$. 
This can happen only if $C_{ij} = 1$ and the pair $ij$ is at risk of transmission at time $t_{ij}$. 
Contact intervals can be right-censored by the end of infectiousness in $i$, by the infection of $j$ from a source other than $i$, or by the end of observation. 
For $i \neq 0$, let $\mathcal{I}_i(t) = \indicator_{t - t_i - \varepsilon_i \in (0, \iota_i]}$ indicate whether $i$ remains infectious at time $t$, and let $\mathcal{I}_0(t)$ indicate whether external infectious contact is possible at time $t$.
Let $\mathcal{S}_{j}(t) = \indicator_{t \leq t_j}$ indicate whether $j$ remains susceptible at time $t$, and let $\mathcal{O}(t) = \indicator_{t \leq T}$ indicate whether observation is ongoing at time $t$. 
Since $\mathcal{I}_i(t)$, $\mathcal{S}_j(t)$, and $\mathcal{O}(t)$ are right-continuous, the processes
\begin{equation}
    Y_{ij}(t) = C_{ij} \mathcal{I}_i(t) \mathcal{S}_j(t) \mathcal{O}(t)
    \label{eq:Yij}
\end{equation}
is right-continuous and equals one when there is a risk of an observed infectious contact from $i$ to $j$ at time $t$. 
The assumptions made in the stochastic S(E)IR model above ensure independent censoring of $\tau_{ij}$ and $t_{0j}$.

\section{Methods}
\label{methods}
Although any parametric failure time distribution could be used in this model, we focus on the following three because they have simple closed-form survival and hazard functions: the exponential distribution with rate $\lambda$, the Weibull distribution with rate $\lambda$ and shape $\gamma$, and the log-logistic distribution with rate $\lambda$ and shape $\gamma$. 
The internal and external transmission models can use the same failure time distribution or different distributions. 
Let the parameters of the internal failure time distribution be $(\lambda_\text{int}, \gamma_\text{int})$ and the parameters of the external distribution be $(\lambda_\text{ext}, \gamma_\text{ext})$, where the shape parameters are omitted for the exponential distribution.

The internal and external transmission models generally work on different time scales (i.e., with different time origins). 
In a pair $ij$ with $i \neq 0$, the time origin is the onset of infectiousness in $i$, which can differ from pair to pair. 
A pair $0j$ is at risk of transmission when $j$ is susceptible and external infectious contact is possible. 
Typically, a common time origin will be specified for all external pairs in a single population under observation.

\subsection{Internal and external rate parameters}
When $i \neq 0$, the rate parameter of the contact interval distribution in the pair $ij$ is
\begin{equation}
    \lambda_{ij} = e^{\beta_\text{int}^\top X_{ij}} \lambda_0,
    \label{eq:internal}
\end{equation}
where $\beta_\text{int}$ is an unknown coefficient vector, $\lambda_0$ is a baseline rate parameter, and $X_{ij}$ is a covariate vector that can include infectiousness covariates for $i$, susceptibility covariates for $j$, and pairwise covariates. 
Each component of $\beta_\text{int}$ is the log rate ratio for a one-unit increase in the corresponding covariate while holding all others constant. 
Because $\lambda_0 > 0$, it will be estimated using an intercept $\ln \lambda_0$.
This model is equivalent to an AFT model where $\exp(-\beta_\text{int}^\text{T} X_{ij})$ is the acceleration factor~\citep{KalbfleischPrentice}. 
We prefer to define the model in terms of rate ratios because the rate ratio is a more common measure of association in epidemiology~\citep{morgenstern1980measures, rothman2008modern}.

The rate parameter for the external contact interval for individual $j$ is
\begin{equation}
  \lambda_{0j} = e^{\beta_\text{ext}^\top X_{0j}} \mu_0,
\end{equation}
where $\beta_\text{ext}$ is an unknown coefficient vector, and $\mu_0$ is the baseline external rate parameter, and $X_{0j}$ is a covariate vector that can include susceptibility covariates for $j$ and environmental or community covariates. 
Like $\beta_\text{int}$, the components of $\beta_\text{ext}$ are log rate ratios and estimation of $\mu_0$ will be done using an intercept $\ln \mu_0$.

There may be overlap between the internal coefficient vector $\beta_\text{int}$ and the external coefficient vector $\beta_\text{ext}$. 
For example, vaccination status could affect the rate parameters for both models. To handle this, we parameterize the combined model as
\begin{equation}
    \lambda_{ij} = e^{\beta^\top X_{ij}} \lambda_0^{1 - \indicator_{i = 0}} \mu_0^{\indicator_{i = 0}}
    \label{eq:lambdaij}
\end{equation}
where the coefficient vector $\beta$ includes coefficients unique to the internal model, coefficients unique to the external model, and shared coefficients. 
The components of $X_{ij}$ used only in the internal model are set to zero when $i = 0$, and the components of $X_{ij}$ used only in the external model are set to zero when $i \neq 0$. 
The distinction between internal and external rows in the data set is maintained using an \emph{external pair indicator} $\zeta = \indicator_{i = 0}$. 
If a covariate in $X_{ij}$ is shared by the internal and external transmission models, it can be allowed to have different coefficients in the two models by including an interaction term with $\zeta$. 
We call these \emph{external interaction terms}.
The parameter vector $X_{ij}$ can include time-varying covariates, which are handled in the same way as in standard survival analysis.

\subsection{Maximum likelihood estimation}
The likelihood and its score process can be derived in a manner similar to that of Kenah\citep{kenah2011contact}. 
Let $\theta$ be a coefficient vector containing the log rate ratios $\beta$, the log baseline rate parameters $\ln \lambda_0$ and $\ln \mu_0$, and the log shape parameters $\ln \gamma_\text{int}$ and $\ln \gamma_\text{ext}$ as needed. 
Let $h_{ij}(t, \theta)$ and $S_{ij}(t, \theta)$ be the hazard and survival functions for the contact interval distribution with rate $\lambda_{ij}$ from equation~\eqref{eq:lambdaij}.
The parametric family may be different for $i \neq 0$ and $i = 0$, which is implemented using the external row indicator $\zeta$. 
Let $\theta_0$ denote the true value of $\theta$. 

\subsubsection{Who-infected-whom observed}
Let $\mathcal{N}_{ij}(t) = \indicator_{t\geq t_{ij}}$ count the first infectious contact from $i$ to $j$.
Assume $j$ is susceptible at time $t = 0$, so $\mathcal{N}_{ij}(0) = 0$. Then $\mathcal{M}_{ij}(t, \theta_0)$ is a mean-zero martingale, where 
\begin{equation}
    \mathcal{M}_{ij}(t, \theta) = \mathcal{N}_{ij}(t) - \int_0^t h_{ij}(u - t_i - \varepsilon_i, \theta) C_{ij} \mathcal{I}_i(u) \dif u
\end{equation}
and we let $t_0 = \varepsilon_0 = 0$.
We observe infectious contacts from $i$ to $j$ only while $j$ is still susceptible and the pair $ij$ is under observation, which gives us the observed counting process
\begin{equation}
    N_{ij}(t) = \int_0^t Y_{ij}(u) \dif \mathcal{N}_{ij}(u).
\end{equation}
Similarly, let
\begin{equation}
    M_{ij}(t, \theta) = \int_0^t Y_{ij}(u) \dif \mathcal{M}_{ij}(u, \theta).
\end{equation}
Then $M_{ij}(t, \theta_0)$ is a mean-zero martingale because it is the integral of a predictable process with respect to the mean-zero martingale $\mathcal{M}_{ij}(u, \theta_0)$.

When we observe infectious contacts from $i$ to $j$ between time $0$ and time $T$, we get the log likelihood 
\begin{equation}
    \ell^*_{ij}(\theta) = \int_0^T \ln h_{ij}(u - t_i - \varepsilon_i, \theta) \dif N_{ij}(u)
    - \int_0^T h_{ij}(u - t_i - \varepsilon_i, \theta) Y_{ij}(u) \dif u.
    \label{eq:l*ij}
\end{equation}
This is a standard log likelihood for right-censored and left-truncated data: 
The first term is a log hazard if $i$ infects $j$ and zero otherwise, and the second term is the negative cumulative hazard of infectious contact. 
The score process is
\begin{equation}
    U^*_{ij}(t, \theta)
    = \int_0^t \left(\frac{\partial}{\partial\theta} \ln h_{ij}(u - t_i - \varepsilon_i, \theta)\right) \dif M_{ij}(u, \theta),
    \label{U*ij}
\end{equation}
which is a mean-zero martingale when $\theta = \theta_0$.

Now fix $j$.
If we observe all pairs $ij$ from time $0$ until time $T$, the log likelihood is 
\begin{equation}
    \ell^*_{\bigcdot j}(\theta) = \sum_{i: i\neq j} \ell^*_{ij}(\theta)
\end{equation}
with score process
\begin{equation}
    U^*_{\bigcdot j}(t, \theta) = \sum_{i: i\neq j} U^*_{ij}(t, \theta),
\end{equation}
which is a mean-zero martingale because it is a sum of mean-zero martingales.  

When we observe who-infected-whom, the log likelihood is $\ell^*(\theta) = \sum_{j=1}^n \ell^*_{\bigcdot j}(\theta)$ and its score process is $U^*(t, \theta) = \sum_{j=1}^n U^*_{\bigcdot j}(t, \theta)$.
Because it is a sum of mean-zero martingales, $U^*(t, \theta_0)$ is also a mean-zero martingale.  
Differentiating $\ell^*(\theta)$ twice, evaluating at $\theta_0$, and taking expectations yields
\begin{equation}
    \E\biggl[-\frac{\partial^2}{\partial \theta^2} \, \ell^*(\theta)
    \,\Big\vert_{\theta = \theta_0}\biggr] 
    = \E\bigl[\big\langle U^*(\theta_0)\big\rangle(T)\bigr],
\end{equation}
where $\big\langle U(\theta_0)\big\rangle(\tau)$ is the predictable variation process of $U(\tau, \theta_0)$.

\subsubsection{Who-infected-whom not observed} 
When who-infected-whom is not observed, we cannot see each $N_{ij}(t)$. 
Instead, we see $N_{\bigcdot j}(t) = \sum_{i\neq j} N_{ij}(t)$. 
The total hazard of infectious contact with $j$ at time $t$ is
\begin{equation}
    h_{\bigcdot j}(t, \theta)
    = \sum_{i: i\neq j} h_{ij}(t - t_i - \varepsilon_i, \theta) C_{ij} \mathcal{I}_i(t),
    \label{eq:lambdaj}
\end{equation}
so the process 
\begin{equation}
    M_{\bigcdot j}(t, \theta)
    = N_{\bigcdot j}(t) - \int_0^t h_{\bigcdot j}(u, \theta) \mathcal{S}_j(u) \mathcal{O}(t) \dif u
    = \sum_{i\neq j} M_{ij}(t, \theta)
\end{equation}
is a mean-zero martingale when $\theta = \theta_0$.  
When $j$ is observed from time $0$ to time $T$, the log likelihood is
\begin{equation}
    \ell_{\bigcdot j}(\theta) = \int_0^T \ln h_{\bigcdot j}(u, \theta) \dif N_{\bigcdot j}(u) - \int_0^T h_{\bigcdot j}(u, \theta)S_j(u) \dif u
    \label{eq:lj}
\end{equation}
and its score process is
\begin{equation}
    U_{\bigcdot j}(t, \theta) = \int_0^t \left[\frac{\partial}{\partial\theta}\ln h_{\bigcdot j}(u, \theta)\right] \dif M_{\bigcdot j}(u, \theta),
    \label{Uj}
\end{equation}
which is a mean-zero martingale when $\theta = \theta_0$.  

The complete-data log likelihood when we do not observe who-infected-whom is $\ell(\theta) = \sum_{j=1}^n \ell_{\bigcdot j}(\theta)$ with score process $U(t, \theta) = \sum_{j=1}^n U_{\bigcdot j}(t, \theta)$. 
Because it is a sum of mean-zero martingales, $U(t, \theta_0)$ is also a mean-zero martingale. 
Differentiating $\ell(\theta)$ twice, evaluating at $\theta_0$, and taking expectations yields
\begin{equation}
    \E\biggl[-\frac{\partial^2}{\partial \theta^2} \, \ell(\theta) 
    \,\Big\vert_{\theta = \theta_0}\biggr] 
    = \E\bigl[\big\langle U(\theta_0)\big\rangle(T)\bigr],
\end{equation}
where $\big\langle U(\theta)\big\rangle(\tau)$ is the predictable variation process of $U(\tau, \theta)$. 

\subsubsection{Pairwise asymptotics} 
The arguments above establish the consistency and asymptotic normality of the maximum likelihood estimator $\hat{\theta}$ as the number of observed infections $m \rightarrow \infty$ as long as the rate of increase in the number of susceptibles at risk of infection is at least as fast as the rate of increase in the number of pairs at risk of transmission~\citep{kenah2011contact, kenah2015semiparametric}. 

\section{Simulations}
The proposed pairwise AFT regression model was tested through $2{,}000$ network-based simulations for each of two different baseline internal contact interval distributions: exponential($\lambda = \ln(-\ln 0.8)$) and log-logistic($\gamma = 2$, $\lambda = 0.5$).
The infectious period was fixed to one time unit, so the household secondary attack risk was $0.2$ in a pair with both covariates equal to zero. 
In all simulations, the external contact interval distribution was exponential with rate $\lambda_\text{ext} = 0.5 \ln(-\ln 0.8)$. 

In each simulation, we generated an undirected network representing $300$ households of size five. 
Each household was a complete graph of size five, and the households were not connected to each other. 
Once a household member was infected, other members of the household could be infected by transmission within the household or by an external source. 
Each epidemic was followed until $500$ infections occurred, which guaranteed at least $200$ infections in individuals who were not primary cases (i.e., the first case in a household).

Each individual $i$ was assigned an independent Bernoulli(0.5) covariate $X_i$. 
The rate parameter for the contact interval distribution in the pair $ij$ was 
\begin{equation}
  \lambda_{ij} = \exp\Big(\beta_\text{inf} X_i + \beta_\text{sus} X_j + \mathbb{I}_{i \neq 0} \ln \lambda_0 + \mathbb{I}_{i = 0} \ln \mu_0\Big), 
\end{equation}
where we set $X_0 = 0$. 
For each simulation, the true values of $\beta_\text{inf}$ and $\beta_\text{sus}$ were independent samples from a uniform($-1, 1$) distribution. 

In each simulation, we analyzed data sets under four different epidemiologic study designs. 
Analysis of within-household transmission is the same for all study designs, but they differ in their inclusion of person-time from individuals at risk of external infectious contact (i.e., pairs $0j$).
The first two study designs are ``valid'' in the sense that their inclusion of pair-time at risk of transmission includes external sources and is predictable with respect to the observed data so it will not generate immortal time bias.
The valid study designs are:
\begin{description}
  \item{\textbf{Complete cohort:}} Follow-up for all $2{,}500$ individuals starts at time zero, which is the time origin for external infectious contact intervals. 
  \item{\textbf{Contact tracing (CT) with delayed entry:}} Follow-up of each individual begins at the infection time of the index case in his or her household. 
    Time at risk of external infectious contact prior to the start of follow-up is left-truncated, and individuals in households with no infections are excluded from the study.
\end{description}
The second two study designs are ``flawed'' in the sense that their inclusion of pair-time at risk of transmission either generates immortal time bias or fails to include external sources of infection.
The flawed study designs are:
\begin{description}
  \item{\textbf{CT without delayed entry:}} Follow-up of all members of households where at least one infection occurs starts retroactively at time zero. 
    Individuals in households with no infections are excluded from the study. 
  \item{\textbf{Ignoring external infection:}} All pairs $0j$ are excluded from the study. 
    This is equivalent to assuming that, in each household, all infections after the primary case are caused by within-household transmission (but not necessarily by the primary case itself).
\end{description}
Under each of the four study designs, data were analyzed both with and without observation of who-infected-whom. 
In all eight analyses of each simulation, we obtained maximum likelihood point estimates of $\beta_\text{inf}$, $\beta_\text{sus}$, $\ln \lambda_0$, $\ln \gamma_\text{int}$, and $\ln \mu_0$. 
For all parameters, we calculated 95\% Wald confidence intervals. 
All regression models used an exponential distribution for external rows and a specified parametric family (exponential, Weibull, or log-logistic) for internal rows.

The epidemic simulations were conducted at the Ohio Supercomputer Center (\url{osc.edu}) using Python version 3.9.12 with SciPy version 1.7.3 (\url{scipy.org}), NetworkX version 2.7.1 (\url{networkx.org}), and pandas version 1.4.2 (\url{pandas.pydata.org}).
The Python simulations use a network-based epidemic simulation script called transtat\_models version 0.2.0 (\url{github.com/ekenah/transtat_models}).
The analysis of data from each simulated epidemic was done using $\mathtt{R}$ version 4.3.1 (\url{cran.r-project.org}) with the packages survival version 3.5-5 and reticulate version 1.31. 
The pairwise AFT models in $\mathtt{R}$ are implemented in the package TranStat version 0.3.7 (\url{github.com/ekenah/TranStat}), which allows pairwise AFT models to be specified using standard $\mathtt{R}$ model syntax. 
All of these software packages are free and open-source. 
The Python and $\mathtt{R}$ code for the simulations, the simulation data, and the $\mathtt{R}$ code for the simulation data analysis are available in the Supplementary Material.

\subsection{Simulation results}
\begin{figure}
  \includegraphics[width=\textwidth]{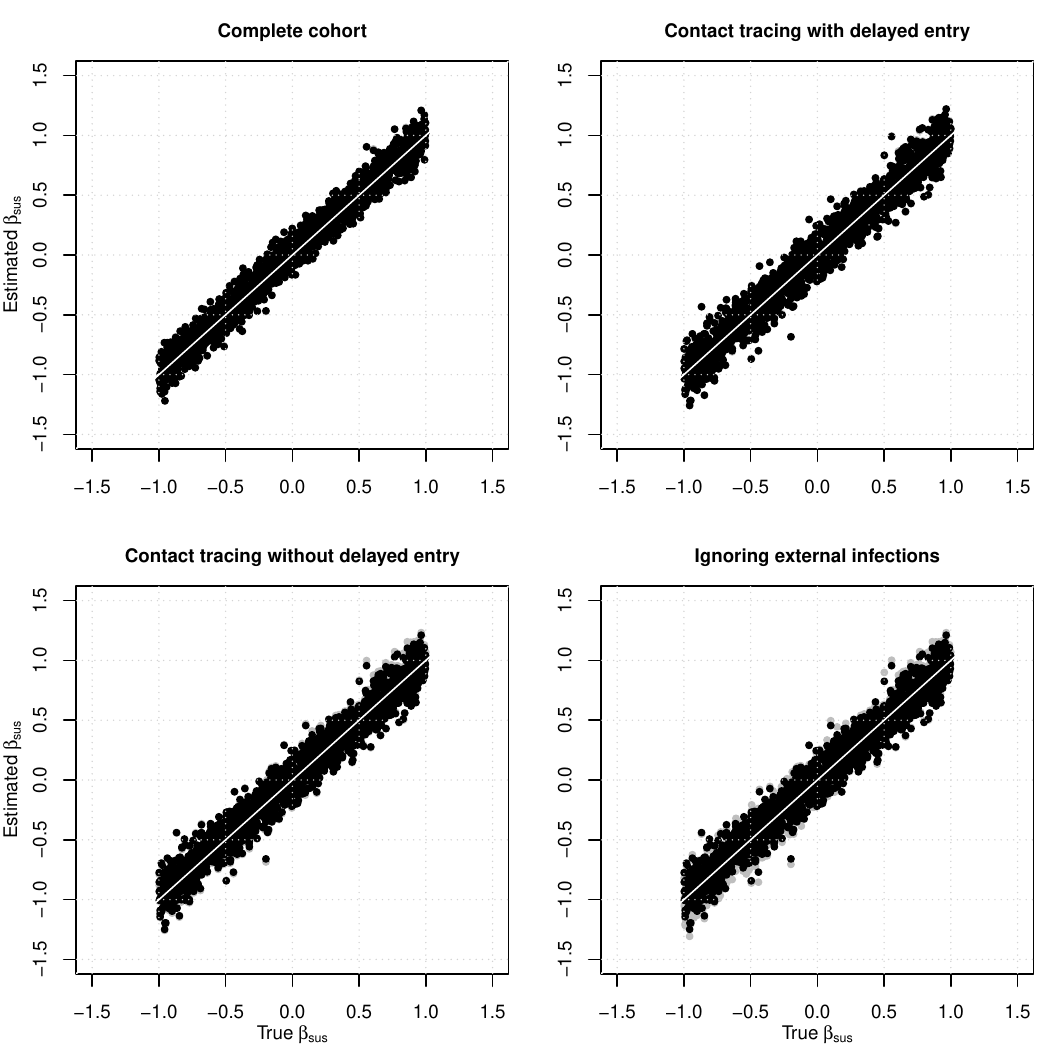}
  \caption{
    Estimated $\beta_\text{sus}$ versus true values for exponential pairwise AFT models fit to simulated data generated with exponential internal contact interval distributions.
    Black dots represent analyses where who-infected-whom was observed, and gray dots represent analyses where who-infected-whom was not observed.
    The line of equality is indicated in white.
  }
  \label{fig:betasus_exp}
\end{figure}

\begin{figure}
  \includegraphics[width=\textwidth]{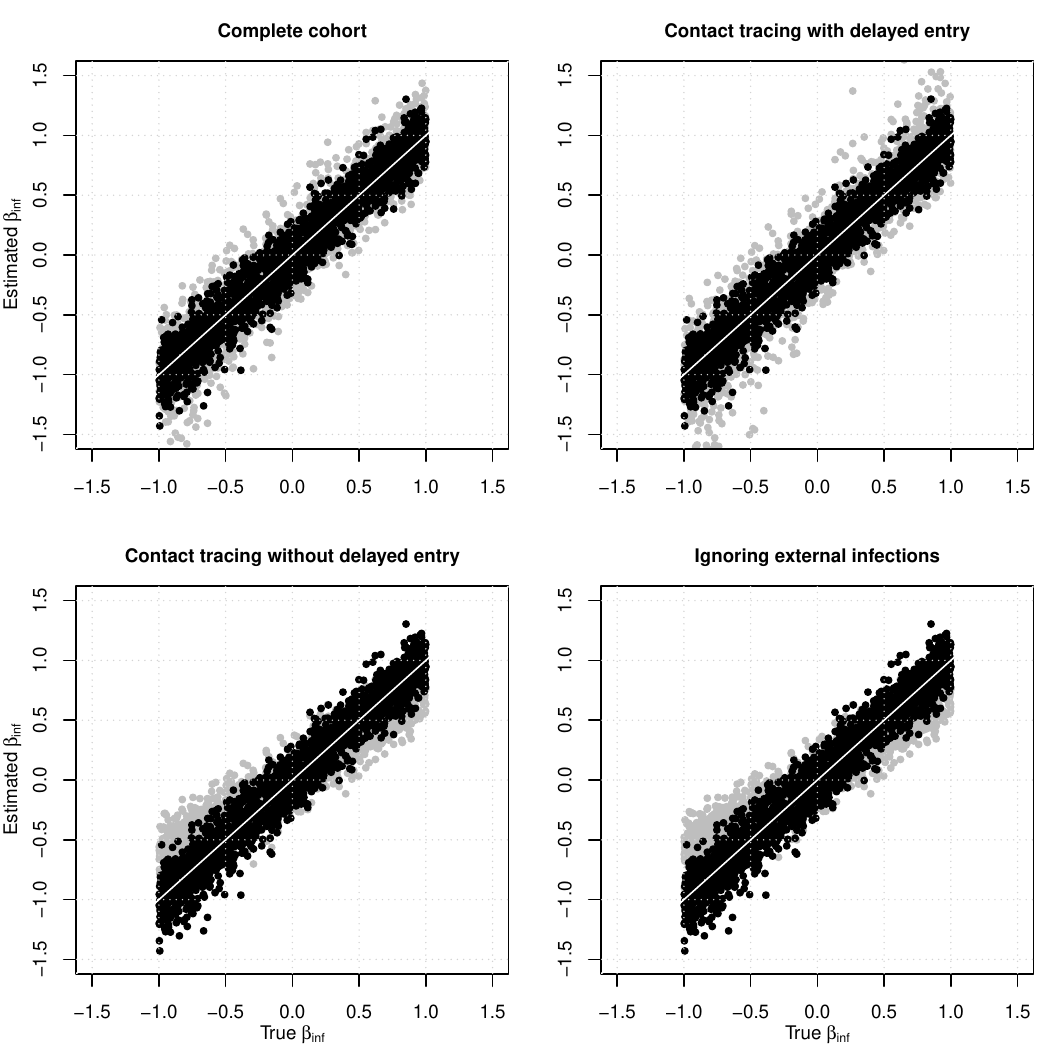}
  \caption{
    Estimated $\beta_\text{inf}$ versus true values for exponential pairwise AFT models fit to simulated data generated with exponential internal contact interval distributions.
    Black dots represent analyses where who-infected-whom was observed, and gray dots represent analyses where who-infected-whom was not observed.
    The line of equality is indicated in white.
  }
  \label{fig:betainf_exp}
\end{figure}

Figure~\ref{fig:betasus_exp} shows scatterplots of the true and estimated $\beta_\text{sus}$ for an exponential pairwise AFT model fit to data generated with exponential internal contact interval distributions.
The gray dots are almost completely hidden behind the black dots, which indicates that observation of who-infected-whom makes little difference to estimation of $\beta_\text{sus}$.
The scatterplots follow the line of equality for all four study designs, so $\beta_\text{sus}$ was estimated with no apparent bias even under the flawed study designs.
Figure~\ref{fig:betainf_exp} show the true and estimated $\beta_\text{inf}$ from the same simulations.
The gray dots show higher variance than the black dots, so observation of who-infected-whom increases the precision of $\beta_\text{inf}$ estimates.
When who-infected-whom is observed, $\beta_\text{inf}$ is estimated with no visible bias under all four study designs.
When who-infected-whom is not observed, there is no visible bias in the valid study designs (top) but there is visible bias toward $\beta_\text{inf} = 0$ in the flawed study designs (bottom).
Roughly speaking, estimation of $\beta_\text{sus}$ depends mostly on who was infected, which is known whether or not we observe who-infected whom.
Estimation of $\beta_\text{inf}$ depends on who caused the infections, about which we have more information if we observe who-infected-whom.

\begin{figure}
  \includegraphics[width=\textwidth]{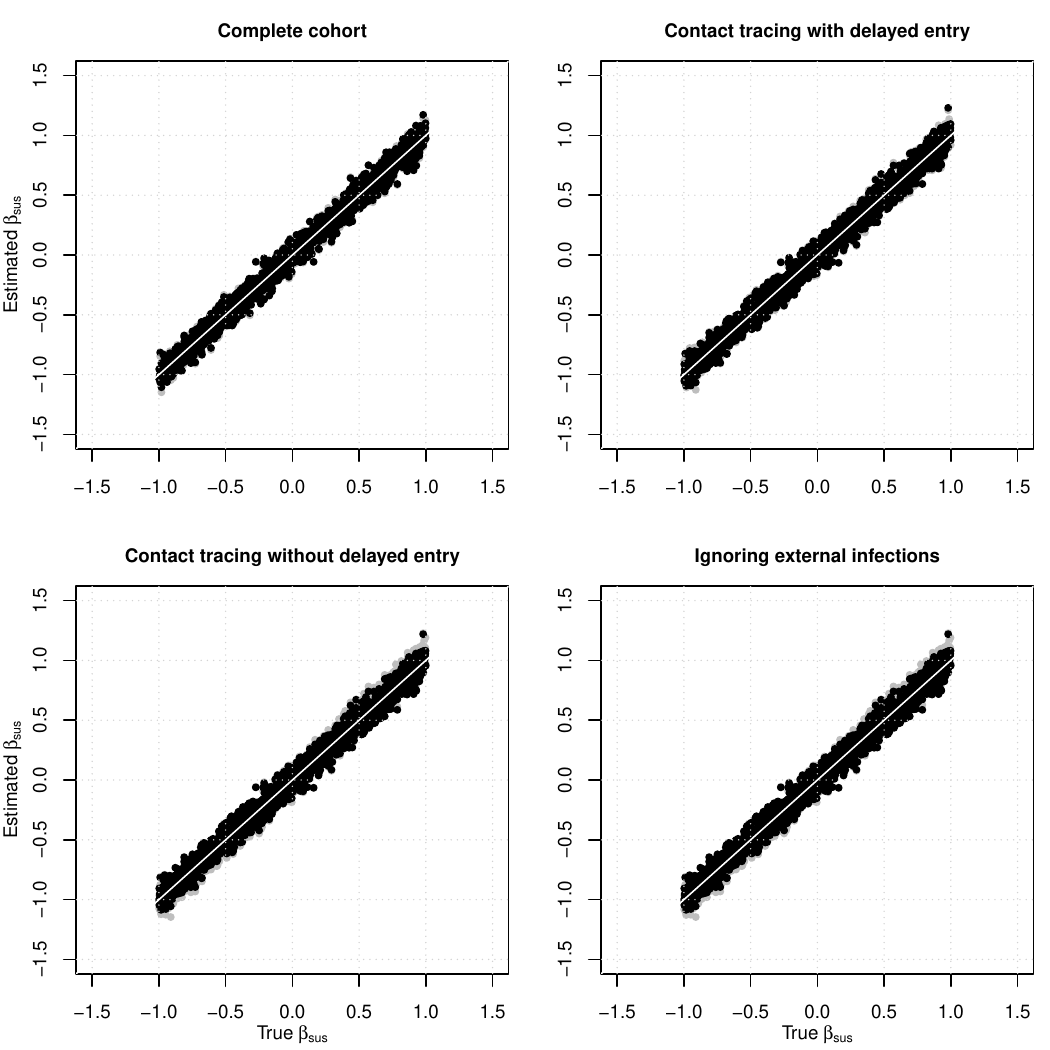}
  \caption{
    Estimated $\beta_\text{sus}$ versus true values for log-logistic pairwise AFT models fit to simulated data generated with log-logistic internal contact interval distributions.
    Black dots represent analyses where who-infected-whom was observed, and gray dots represent analyses where who-infected-whom was not observed.
    The line of equality is indicated in white.
  } 
  \label{fig:betasus_llog}
\end{figure}

\begin{figure}
  \includegraphics[width=\textwidth]{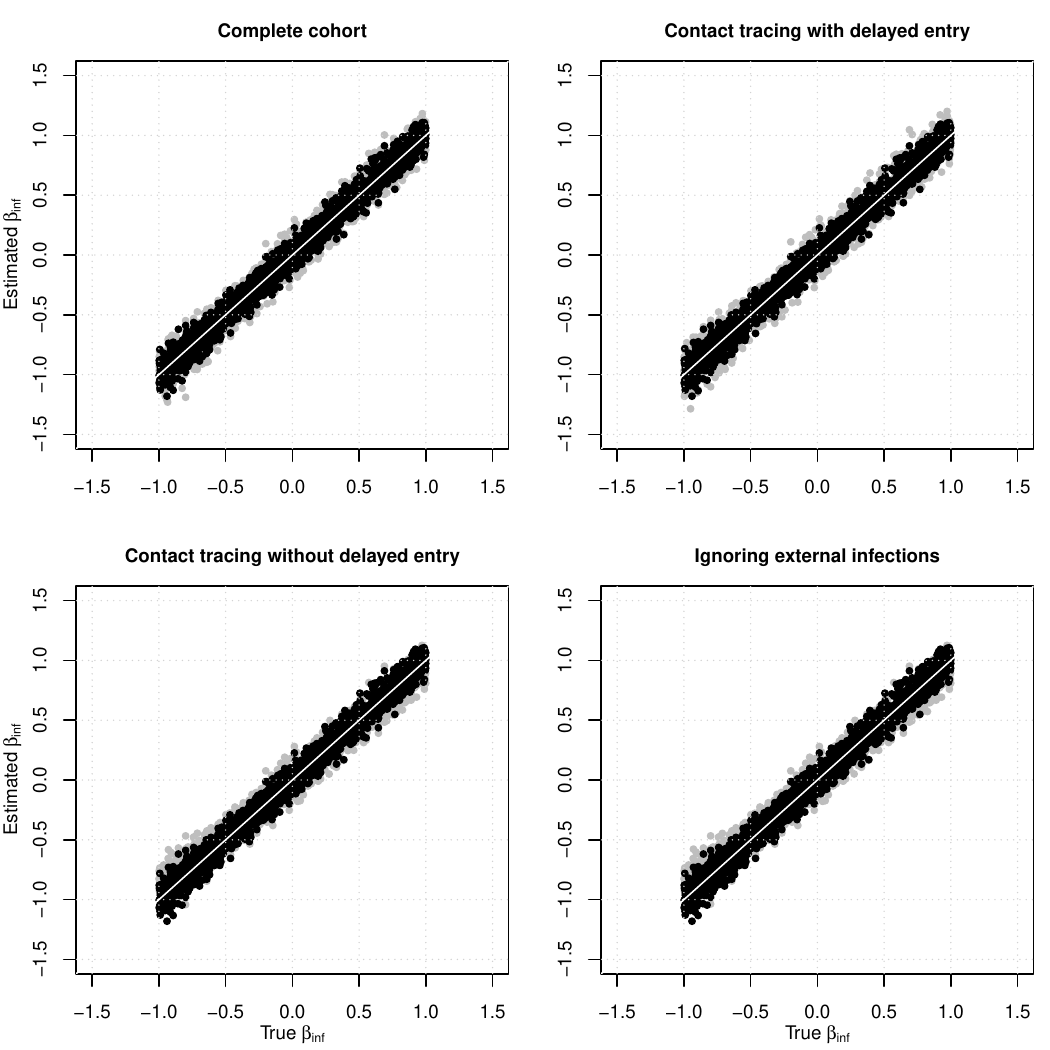}
  \caption{
    Estimated $\beta_\text{inf}$ versus true values for log-logistic pairwise AFT models fit to simulated data generated with log-logistic internal contact interval distributions.
    Black dots represent analyses where who-infected-whom was observed, and gray dots represent analyses where who-infected-whom was not observed.
    The line of equality is indicated in white.
  } 
  \label{fig:betainf_llog}
\end{figure}

Figure~\ref{fig:betasus_llog} and Figure~\ref{fig:betainf_llog} show scatterplots of the true and the estimated $\beta_\text{sus}$ and $\beta_\text{inf}$ for a log-logistic pairwise AFT model fit to data generated with log-logistic internal contact interval distributions. 
Estimation of $\beta_\text{sus}$ has is almost unaffected by observation of who-infected-whom, and it appears unbiased even under the flawed epidemiologic study designs.
Estimation of $\beta_\text{inf}$ has lower variance when who-infected-whom is observed, but this difference is smaller than it was for the correctly-specified exponential pairwise AFT models in Figure~\ref{fig:betainf_exp}.
When who-infected-whom is observed, $\beta_\text{inf}$ is estimated with no visible bias under all four study designs.
When who-infected-whom is not observed, there is no visible in the valid study designs (top) and a slight bias toward $\beta_\text{inf} = 0$ in the flawed study designs (bottom).
Overall, we see the same pattern as in Figures~\ref{fig:betasus_exp} and~\ref{fig:betainf_exp}.
Estimation of $\beta_\text{inf}$ is more sensitive to observation of who-infected-whom and to epidemiologic study design than estimation of $\beta_\text{sus}$.

For all parameters, Table~\ref{tab:coverage} shows the coverage probabilities for Wald 95\% confidence intervals for correctly-specified exponential and log-logistic pairwise AFT models.
The valid study designs produce nominal coverage probabilities for all parameters whether or not who-infected-whom is observed.
When who-infected-whom is observed, the flawed study designs produce near-nominal coverage probabilities for $\beta_\text{inf}$, $\beta_\text{sus}$, and the log-logistic $\ln \gamma_0$ but very low coverage probabilities (or no estimates at all) for $\ln \mu_0$.
When who-infected-whom is not observed, the flawed study designs produce low coverage probabilities for all parameters except $\beta_\text{sus}$, but the correctly-specified log-logistic model performed substantially better than the correctly-specified exponential model. 

\begin{figure}
  \includegraphics[width=\textwidth]{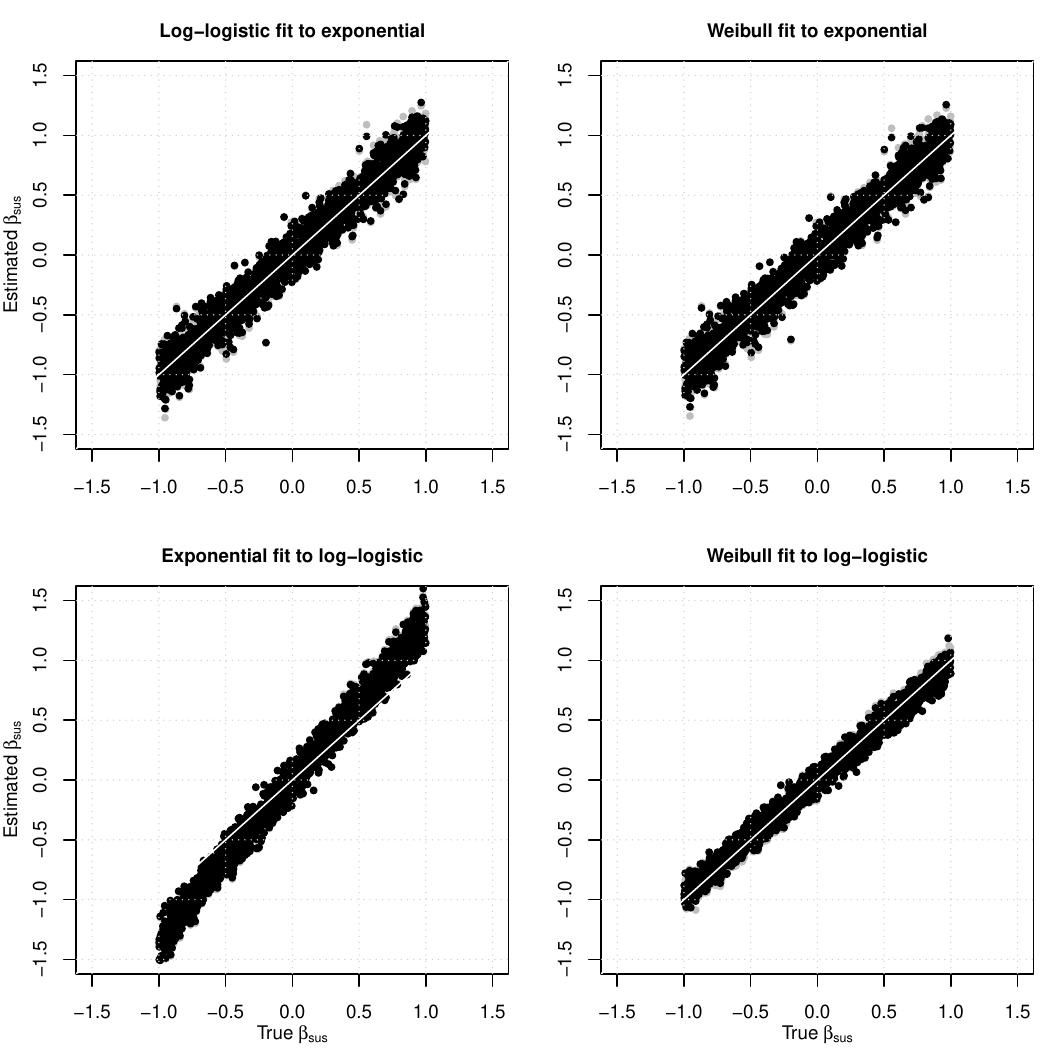}
  \caption{
    Estimated $\beta_\text{sus}$ versus true values for log-logistic and Weibull pairwise AFT models fit to data generated with exponential internal contact interval distributions (top) and for exponential and Weibull pairwise AFT models fit to data generated with log-logistic internal contact interval distributions (bottom).
    Black dots represent analyses where who-infected-whom was observed, gray dots represent analyses where who-infected-whom was not observed, and the line of equality is indicated in white.
  } 
  \label{fig:betasus_mis}
\end{figure}

\begin{figure}
  \includegraphics[width=\textwidth]{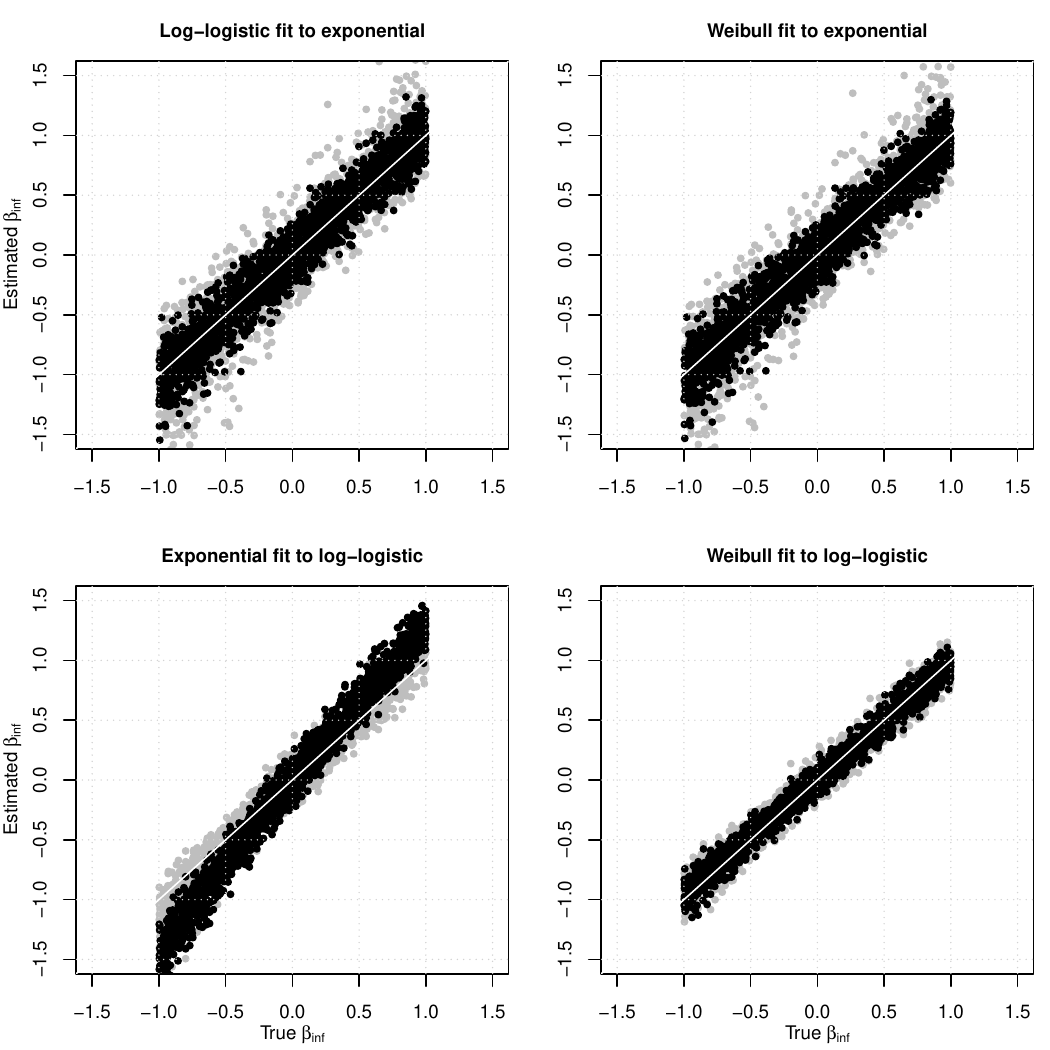}
  \caption{
    Estimated $\beta_\text{inf}$ versus true values for log-logistic and Weibull pairwise AFT models fit to data generated with exponential internal contact interval distributions (top) and for exponential and Weibull pairwise AFT models fit to data generated with log-logistic internal contact interval distributions (bottom).
    Black dots represent analyses where who-infected-whom was observed, gray dots represent analyses where who-infected-whom was not observed, and the line of equality is indicated in white.
  } 
  \label{fig:betainf_mis}
\end{figure}

We also analyzed each set of simulation results using pairwise AFT models with internal contact interval distributions from a different parametric family than the one used to generate the data.
The top of Figure~\ref{fig:betasus_mis} shows estimates of $\beta_\text{sus}$ from log-logistic and Weibull pairwise AFT models fit to data generated with exponential contact intervals.
As before, observation of who-infected-whom had almost no effect on these estimates.
Both models estimated $\beta_\text{sus}$ with little or no bias, and their estimates are remarkably similar.
The bottom of Figure~\ref{fig:betasus_mis} shows estimates of $\beta_\text{sus}$ from exponential and Weibull pairwise AFT models fit to data generated with log-logistic contact intervals.
The exponential model produced estimates that are clearly biased away from $\beta_\text{sus} = 0$.
The Weibull model produced estimates that are nearly unbiased.
The top of Figure~\ref{fig:betainf_mis} shows estimates of $\beta_\text{inf}$ from log-logistic and Weibull pairwise AFT models fit to data generated with exponential contact intervals.
As before, the estimates have lower variance when who-infected-whom is observed.
Both models estimated $\beta_\text{inf}$ with little or no bias whether or no who-infected-whom was observed, and their estimates are remarkably similar in both cases.
The bottom of Figure~\ref{fig:betainf_mis} shows estimates of $\beta_\text{inf}$ from exponential and Weibull pairwise AFT models fit to data generated with log-logistic contact intervals.
The exponential model produced estimates that are clearly biased, especially when who-infected-whom is observed.
The Weibull model produced estimates that are nearly unbiased whether or not who-infected-whom was observed.
For estimation of rate ratios, it appears to be more important that a pairwise AFT model have sufficient flexibility than it is to choose exactly the right parametric family.
The log-logistic pairwise AFT model produced accurate rate ratio estimates when fit to data generated with exponential or Weibull contact intervals, neither of which is a special case of the log-logistic distribution.
Similarly, the Weibull pairwise AFT model produced accurate rate ratio estimates for data generated with log-logistic contact intervals.
In contrast, the exponential model produced severely biased estimates of rate ratios when fit to data generated with Weibull or log-logistic contact intervals.

Table~\ref{tab:coverage_exp} shows coverage probabilities from pairwise AFT models with Weibull and log-logistic internal contact intervals analyzing simulated data generated with exponential internal contact intervals.
Because the exponential distribution is a special case of the Weibull distribution, the Weibull model produces results similar to the correctly-specified pairwise AFT models in Table~\ref{tab:coverage} (except for $\ln \mu_0$ when who-infected-whom is not observed). 
In particular, the coverage probability for the Weibull shape parameter (with true value $\gamma = 1$) is near-nominal for all study designs.
Because the exponential distribution is not a special case of the log-logistic distribution, the log-logistic model is misspecified.
There is no true value for the log-logistic log shape $\ln \gamma$, and $\ln \lambda_0$ has a different interpretation in the two models.
Nonetheless, the coverage probabilities for $\beta_\text{sus}$ are near-nominal for all study designs, and the coverage probabilities for $\beta_\text{inf}$ are near-nominal for all study designs when who-infected-whom is observed and for the valid study designs when who-infected-whom is not observed.

Table~\ref{tab:coverage_llog} shows coverage probabilities from pairwise AFT models with exponential and Weibull internal contact intervals analyzing simulated data generated with log-logistic internal contact intervals.
Because the log-logistic distribution is not a special case of the Weibull distribution, both of these models are misspecified.
The exponential model produced low coverage probabilities when who-infected-whom was observed and a mixture of low and abnormally high coverage probabilities (suggesting high variance) when who-infected-whom was not observed.
The Weibull model produced near-nominal coverage probabilities for $\beta_\text{sus}$ and $\beta_\text{inf}$ under the valid study designs whether or not who-infected-whom was observed.
Under the flawed study designs, it produced near-nominal coverage probabilities for $\beta_\text{sus}$ but substantially lower coverage probabilites for $\beta_\text{inf}$.
The Weibull model produced consistently low coverage probabilities for $\ln \lambda_0$ and $\ln \gamma_\text{int}$, which have different interpretations in the log-logistic and Weibull distributions.

Because so little is known about infectiousness profiles of communicable diseases, a parametric family for the contact interval distribution will usually need be chosen empirically.
Table~\ref{tab:aic} shows that the correctly-specified model usually had the lowest Akaike Information Criterion~\citep[AIC;][]{akaike1974new}, suggesting that the choice of a parametric model can be guided by model fit.

\section{Los Angeles County influenza A(H1N1) data}
To give an example of pairwise AFT modeling of infectious disease transmission data, we analyze influenza A (H1N1) household surveillance data collected by the Los Angeles County Department of Public Health (LACDPH) in April and May, 2009. 
The data was collected using the following protocol~\citep{Sugimoto2011}:
\begin{enumerate}
  \item Between April 14 and May 18, nasopharyngeal swabs and aspirates were taken from individuals who reported to the LACDPH or other local health care providers with acute febrile respiratory illness (AFRI), defined as a fever $\geq 37.8^\circ \text{C}$ plus at least one of cough, sore throat, or rhinorrhea (runny nose). 
    These specimens were tested for influenza using reverse transcriptase polymerase chain reaction (RT-PCR). 
  \item Patients whose specimens tested positive for pandemic influenza A (H1N1) or for influenza A of undetermined subtype were invited to participate in a phone interview. 
    These interviews used a standard questionnaire developed by the LACDPH to to collect information about his or her household contacts, including sex, age, and antiviral prophylaxis use. 
    For index cases under 18 years of age, an adult proxy was interviewed. 
  \item The initial interview and, when necessary, a follow-up interview were used to obtain the symptom onset dates of AFRI episodes in the household up to 14 days after the symptom onset date of the index case. 
    All interviews were completed between April 30 and June 1.
\end{enumerate}

For simplicity, we assume all AFRI episodes among household members were caused by influenza A (H1N1). 
All index cases are assumed to be external infections, and all other household members are assumed to be at risk of infection from both household members and external sources. 
This study design corresponds to contact tracing with delayed entry in the simulation study above.

The primary analysis assumed an incubation period of 2 days, a latent period of 0 days, and an infectious period of 6 days. These natural history assumptions are adapted from~Yang et al.~\citep{YangH1N1}. 
Households were identified upon clinical presentation of an index case, so household members were considered to be at risk of infection from the infection time of the index case (which depends on the assumed incubation period) until 14 days after the infection time of the index case. 
In a sensitivity analysis, we varied the assumed latent and infectious periods.

The covariates used in our analysis were sex ($\mathtt{male} = 1$ for males and $\mathtt{male} = 0$ for females), age category ($\mathtt{adult} = 1$ for ages $\geq 18$ years and $\mathtt{adult} = 0$ otherwise), and antiviral prophylaxis. 
Antiviral prophylaxis was assumed to be initiated on the day following the symptom onset of the index case in each household, so it was handled as a time-dependent covariate. 
Each pair had covariate values for the infectious individual ($\mathtt{male\_inf}$, $\mathtt{adult\_inf}$, and $\mathtt{proph\_inf}$) and for the susceptible individual ($\mathtt{male\_sus}$, $\mathtt{adult\_sus}$, and $\mathtt{proph\_sus}$). 
In external pairs, all infectiousness covariates were set to zero. 
We considered exponential, Weibull, and log-logistic distributions for the internal and external contact interval distributions. 
All models were fit using the Broyden, Fletch, Goldfarb, and Shanno algorithm (\texttt{BFGS} in the \texttt{R} function \texttt{optim}) with starting parameter values taken from an initial fit using exponential internal and external contact intervals.  

Statistical analysis was conducted in $\mathtt{R}$ version 4.3.1 (\url{www.r-project.org}) using TranStat version 0.3.7 (\url{gihub.com/ekenah/TranStat}). 
The data set and analysis code are available in the Supplementary Material.

\subsection{Data analysis results}
The household data collected by the Los Angeles County Department of Public Health included 299 individuals in 58 households. 
There were 99 probable influenza infections, of which 62 were index cases---four households had co-primary cases with symptom onsets on the same day. 
There were three people missing data on sex, four people missing data on age, and 56 people missing data on antiviral prophylaxis. 
The 62 individuals with missing data came from 17 households with 36 infections, of which 19 were index cases. 
Because we assume all household members can infect or be infected by other household members, we excluded the entire household if any of its members was missing data. 
In the complete-cases data set, we have 41 households with 63 infections, of which 43 were index cases.

Using the complete-cases data set, we fit a model with main effects for all six covariates using all nine possible combinations of internal and external contact interval distributions.
All models were fit using the BFGS algorithm for optimization as in the simulations, which is the default in TranStat.
Table~\ref{tab:LAaic} shows the resulting AIC values. 
The three minimum AIC values occur for exponential internal contact intervals.
Among these three, the lowest AIC occurs for log-logistic external contact intervals. 
Using exponential internal contact intervals and log-logistic external contact intervals, we built a model using backwards selection to achieve the minimum AIC.
This removed all covariates except for three: age category for infectiousness ($\mathtt{adult\_inf}$), age category for susceptibility ($\mathtt{adult\_sus}$), and prophylaxis by susceptibles ($\mathtt{proph\_sus}$). 
The AIC of this model was $203.59$.

We then checked for external interaction terms, which allow a covariate to have different coefficients in the internal and external transmission models. 
An external interaction term with $\mathtt{adult\_sus}$ had a p-value of $0.89$ and increased the AIC to $205.57$. 
An external interaction term with $\mathtt{proph\_sus}$ had a p-value of $0.87$ but reduced the AIC to $202.37$.
A joint likelihood ratio test for the main effect and external interaction term for $\mathtt{proph\_sus}$ in this model yielded a p-value of $0.008$, which is consistent with the p-value of $0.012$ for $\mathtt{proph\_sus}$ in the model with no interaction term.
For simplicity, we did not keep this interaction term in the model. 

Our final model is summarized at the top of Table~\ref{tab:coefs}.
The coefficients for covariates are log rate ratios, $\mathtt{intercept} = \ln \lambda_0$ (the log baseline internal rate parameter), $\mathtt{xintercept} = \ln \mu_0$ (the log baseline external rate parameter), and $\mathtt{xlogshape} = \ln \gamma_\text{ext}$ (the external log-logistic shape parameter).
The BFGS algorithm did not find an upper 95\% likelihood ratio confidence limit for $\mathtt{adult\_inf}$, so we calculated this upper limit by refitting the model using the Nelder-Mead algorithm.
Both algorithms produced nearly identical point and interval estimates for all other parameters in the model.

The model suggests that adults were more infectious and less susceptible than children, but the small number of transmission events observed makes these results inconclusive. 
The predicted rate ratio for susceptibility associated with antiviral prophylaxis is $0.34$ $(0.10, 0.76)$, so the model strongly suggests that antiviral prophylaxis in susceptibles reduced their risk of infection. 
We found no clear evidence of differences in infectiousness or susceptibility by sex, and we found no clear evidence of an effect of antiviral prophylaxis on infectiousness. 

Table~\ref{tab:sar} shows the predicted household secondary attack risk (SAR) by the age of the infectious individual, the age of the susceptible individual, and antiviral prophylaxis in the susceptible individual. 
The higher infectiousness and lower susceptibility of adults is apparent, as is the protective effect of antiviral prophylaxis. 
Because the predicted household SAR depends on multiple parameters in the regression model, we used Wald confidence intervals. 

To see how accounting for external sources of infection affected our analysis, we re-fit our final model using only data on infectious-susceptible pairs within households. 
This model is summarized at the bottom of Table~\ref{tab:coefs}. 
The two models give similar results, but accounting for external infection gave us greater statistical power to estimate the effects of age and antiviral prophylaxis. 
Using a two-parameter contact interval distribution did not restore the statistical power lost by ignoring external sources of infection.
Refitting the final model without external rows using Weibull or log-logistic contact intervals instead of exponential contact intervals yielded a p-value of $0.160$ for the coefficient on antiviral prophylaxis in susceptible individuals.

Table~\ref{tab:sensitivity} shows the results of a sensitivity analysis where we varied assumptions about the latent and infectious periods. 
The infectiousness rate ratio for age category and its p-value are highly sensitive to the assumed latent and infectious periods. 
The susceptibility rate ratio for age category and its p-value are somewhat more stable. 
The susceptibility rate ratio for antiviral prophylaxis and its p-value are remarkably stable. 
The rate ratio varies from $0.35$ to $0.41$, and its p-value varies from $0.011$ to $0.026$. 
The loss of statistical power when we fail to account for external sources of infection is consistent throughout the sensitivity analysis.

\section{Discussion}
Our simulation results showed that the pairwise AFT model produces reliable point and interval estimates of parameters governing the internal and external contact interval distributions.
In particular, it produced reliable estimates of rate ratios for infectiousness and susceptibility under valid epidemiologic study designs whether or not who-infected-whom was observed.
When who-infected-whom was observed, these rate ratio estimates were surprisingly robust to flawed epidemiologic study design.
Estimates of rate ratios for infectiousness were more sensitive to both observation of who-infected-whom and epidemiologic study design than estimates of rate ratios for susceptibility.
A sufficiently flexible pairwise AFT model can accurately estimate rate ratios even when it is slightly misspecified (e.g., a Weibull model used for data generated with log-logistic contact intervals).
It is likely that the simulation results would have been even better if we had used likelihood ratio confidence intervals or taken steps to identify a good starting point for maximization of the likelihood.

There are several limitations of the LACDPH household data analysis that point toward future research topics.
In the interest of simplicity, the handling of missing data was crude.
We removed entire households when any member was missing a covariate, and we assumed fixed incubation, latent, and infectious periods to avoid treating these times as missing. 
Bayesian methods with data imputation~\citep{ONeillRoberts1999} would be a more principled way to handle missing data.
With no clear scientific basis for choosing parametric families for the internal and external contact interval distributions, we compared models using the AIC.
Although our simulations suggest this can be a reliable approach when the available parametric families include the true distributions, an extension of the semiparametric model of Kenah~\citep{kenah2015semiparametric} that could handle external sources of infection would not require a choice of parametric families.
Finally, the pairwise AFT model showed occasional numerical instability. 
Although we dealt successfully with this problem ad hoc, it deserves more systematic investigation.

The pairwise AFT model can be viewed as an extension of the longitudinal chain-binomial model~\citep{RampeyLongini1992, BeckerBritton1999} to continuous time. 
Like these models, it accounts for dependent events and for infection from external sources even when who-infected-whom is not observed.
Unlike these models, it allows flexibility in the infectiousness profile without a large number of nuisance parameters, and it can be specified, fit, and interpreted in a manner similar to standard regression models. 
The simulation study showed that it produces accurate point and interval estimates when the epidemiologic study design is valid and the chosen parametric models have the flexibility to mimic the true internal and external contact interval distributions.
The analysis of the LACDPH influenza A (H1N1) household data showed that the model can produce insights relevant to public health and that accounting for external sources of infection is important.
When who-infected-whom is observed (which occurs rarely in practice), estimation of rate ratios for susceptiblity and infectiousness is remarkably robust to flawed epidemiologic study design.
This suggests that pathogen phylogenies, which provide partial information on who-infected-whom, could improve precision and reduce bias~\citep{kenah2016}.
With or without pathogen genome sequences, pairwise AFT regression models can generate detailed insights about the transmission of infectious diseases from longitudinal studies of close contact groups or from contact tracing data.
These insights can help us design efficient and effective public health interventions to control future epidemics.

\section*{Acknowledgements} 
  The authors would like to thank Forrest Crawford (Yale School of Public Health) and Jonathan Sugimoto (Fred Hutchinson Cancer Research Center) for their comments, and we are grateful to the Los Angeles County Department of Public Health (LACDPH) for allowing the use of their data.
  YS was supported by National Institutes of Health (NIH) grant DP2HD09179. 
  YS, ZD, WRKB, and EK were supported by National Institute of Allergy and Infectious Diseases Grant (NIAID) grant R01 AI116770.
  YS and EK were supported by NIAID grant R03 AI124017. 
  EK was supported by National Institute of General Medical Sciences (NIGMS) grant U54 GM111274 and NIAID grant U01 AI169375. 
  WRKB was supported by the President's Postdoctoral Scholars Program at the Ohio State University (OSU), by a Scheme 4 grant (Ref. 42118) from the London Mathematical Society (LMS), and an International Collaboration Fund awarded by the Faculty of Science, University of Nottingham (UoN). 
  The content is solely the responsibility of the authors and does not represent the official views or policies of LACDPH, LMS, NIGMS, NIH, NIAID, OSU, or UoN.

\section*{Supplementary materials}
\begin{description}
  \item{\textbf{pAFTsims.R:}}
    $\mathtt{R}$ script to generate epidemic simulations and fit pairwise AFT models.
  \item{\textbf{pAFTsimulation.py:}}
    Python script for epidemic simulation called by \texttt{pAFTsims.R}.
  \item{\textbf{results.csv:}}
    Comma-separated values (CSV) file with simulation results.
  \item{\textbf{pAFTanalysis.R:}}
    $\mathtt{R}$ script to analyze \texttt{results.csv}, which generates Figures~1-6 and Tables~1-4.
  \item{\textbf{LAdata\_2023-08.csv:}}
    Los Angeles County Department of Public Health household influenza transmission data.
  \item{\textbf{AFT\_LAanalysis.R:}}
    $\mathtt{R}$ script for LACDPH household data analysis, which generates tables 5-8.
\end{description}

\bibliographystyle{chicago}
\bibliography{pairwiseAFT}

\begin{table}[ht]
\centering
\caption{95\% confidence interval coverage probabilities for correctly-specified pairwise AFT models. The flawed study designs are in italics.} 
\label{tab:coverage}
\begingroup\footnotesize
\begin{tabular}{llccccc}
  \toprule
Contact intervals & Study design & $\beta_\text{sus}$ & $\beta_\text{inf}$ & $\ln \lambda_0$ & $\ln \gamma_\text{int}$ & $\ln \mu_0$ \\ 
 \midrule 
\multicolumn{7}{l}{\textbf{Who-infected-whom observed}} \\\multirow{4}{*}{Exponential} & Complete cohort & 0.936 & 0.951 & 0.947 &  & 0.953 \\ 
   & CT with delayed entry & 0.952 & 0.951 & 0.946 &  & 0.958 \\ 
   & \emph{CT without delayed entry} & 0.959 & 0.951 & 0.947 &  & 0.000 \\ 
   & \emph{Ignoring external infection} & 0.954 & 0.950 & 0.943 &  &  \\ 
  \cmidrule{2-7}\multirow{4}{*}{Log-logistic} & Complete cohort & 0.945 & 0.947 & 0.947 & 0.938 & 0.915 \\ 
   & CT with delayed entry & 0.949 & 0.948 & 0.944 & 0.936 & 0.960 \\ 
   & \emph{CT without delayed entry} & 0.950 & 0.949 & 0.943 & 0.936 & 0.000 \\ 
   & \emph{Ignoring external infection} & 0.948 & 0.947 & 0.941 & 0.939 &  \\ 
  \midrule 
\multicolumn{7}{l}{\textbf{Who-infected-whom not observed}} \\\multirow{4}{*}{Exponential} & Complete cohort & 0.937 & 0.955 & 0.952 &  & 0.945 \\ 
   & CT with delayed entry & 0.952 & 0.963 & 0.959 &  & 0.938 \\ 
   & \emph{CT without delayed entry} & 0.952 & 0.785 & 0.341 &  & 0.196 \\ 
   & \emph{Ignoring external infection} & 0.951 & 0.740 & 0.224 &  &  \\ 
  \cmidrule{2-7}\multirow{4}{*}{Log-logistic} & Complete cohort & 0.948 & 0.950 & 0.943 & 0.941 & 0.919 \\ 
   & CT with delayed entry & 0.954 & 0.948 & 0.936 & 0.934 & 0.915 \\ 
   & \emph{CT without delayed entry} & 0.949 & 0.929 & 0.911 & 0.716 & 0.894 \\ 
   & \emph{Ignoring external infection} & 0.950 & 0.928 & 0.911 & 0.711 &  \\ 
   \bottomrule
\end{tabular}
\endgroup
\end{table}

\begin{table}[ht]
\centering
\caption{95\% confidence interval coverage probabilities for pairwise AFT models fit to exponential contact intervals. The flawed study designs are in italics.} 
\label{tab:coverage_exp}
\begingroup\footnotesize
\begin{tabular}{llccccc}
  \toprule
Model & Study design & $\beta_\text{sus}$ & $\beta_\text{inf}$ & $\ln \lambda_0$ & $\ln \gamma_\text{int}$ & $\ln \mu_0$ \\ 
 \midrule 
\multicolumn{7}{l}{\textbf{Who-infected-whom observed}} \\\multirow{4}{*}{Log-logistic} & Complete cohort & 0.934 & 0.948 & 0.748 &  & 0.956 \\ 
   & CT with delayed entry & 0.954 & 0.947 & 0.749 &  & 0.959 \\ 
   & \emph{CT without delayed entry} & 0.957 & 0.947 & 0.745 &  & 0.000 \\ 
   & \emph{Ignoring external infection} & 0.952 & 0.947 & 0.763 &  &  \\ 
  \cmidrule{2-7}\multirow{4}{*}{Weibull} & Complete cohort & 0.933 & 0.945 & 0.942 & 0.946 & 0.954 \\ 
   & CT with delayed entry & 0.951 & 0.944 & 0.939 & 0.950 & 0.958 \\ 
   & \emph{CT without delayed entry} & 0.954 & 0.946 & 0.940 & 0.947 & 0.000 \\ 
   & \emph{Ignoring external infection} & 0.951 & 0.946 & 0.933 & 0.948 &  \\ 
  \midrule 
\multicolumn{7}{l}{\textbf{Who-infected-whom not observed}} \\\multirow{4}{*}{Log-logistic} & Complete cohort & 0.933 & 0.950 & 0.801 &  & 0.837 \\ 
   & CT with delayed entry & 0.954 & 0.963 & 0.817 &  & 0.929 \\ 
   & \emph{CT without delayed entry} & 0.948 & 0.830 & 0.128 &  & 0.195 \\ 
   & \emph{Ignoring external infection} & 0.945 & 0.782 & 0.073 &  &  \\ 
  \cmidrule{2-7}\multirow{4}{*}{Weibull} & Complete cohort & 0.930 & 0.956 & 0.925 & 0.940 & 0.785 \\ 
   & CT with delayed entry & 0.949 & 0.959 & 0.946 & 0.952 & 0.930 \\ 
   & \emph{CT without delayed entry} & 0.950 & 0.793 & 0.583 & 0.949 & 0.196 \\ 
   & \emph{Ignoring external infection} & 0.953 & 0.749 & 0.470 & 0.951 &  \\ 
   \bottomrule
\end{tabular}
\endgroup
\end{table}

\begin{table}[ht]
\centering
\caption{95\% confidence interval coverage probabilities for pairwise AFT models fit to log-logistic contact intervals. The flawed study designs are in italics.} 
\label{tab:coverage_llog}
\begingroup\footnotesize
\begin{tabular}{llccccc}
  \toprule
Model & Study design & $\beta_\text{sus}$ & $\beta_\text{inf}$ & $\ln \lambda_0$ & $\ln \gamma_\text{int}$ & $\ln \mu_0$ \\ 
 \midrule 
\multicolumn{7}{l}{\textbf{Who-infected-whom observed}} \\\multirow{4}{*}{Exponential} & Complete cohort & 0.693 & 0.645 & 0.000 &  & 0.854 \\ 
   & CT with delayed entry & 0.637 & 0.646 & 0.000 &  & 0.928 \\ 
   & \emph{CT without delayed entry} & 0.664 & 0.645 & 0.000 &  & 0.000 \\ 
   & \emph{Ignoring external infection} & 0.624 & 0.645 & 0.000 &  &  \\ 
  \cmidrule{2-7}\multirow{4}{*}{Weibull} & Complete cohort & 0.916 & 0.922 & 0.034 & 0.013 & 0.911 \\ 
   & CT with delayed entry & 0.912 & 0.924 & 0.043 & 0.015 & 0.959 \\ 
   & \emph{CT without delayed entry} & 0.906 & 0.923 & 0.045 & 0.015 & 0.000 \\ 
   & \emph{Ignoring external infection} & 0.913 & 0.923 & 0.044 & 0.016 &  \\ 
  \midrule 
\multicolumn{7}{l}{\textbf{Who-infected-whom not observed}} \\\multirow{4}{*}{Exponential} & Complete cohort & 0.687 & 0.894 & 0.000 &  & 0.860 \\ 
   & CT with delayed entry & 0.621 & 0.972 & 0.000 &  & 1.000 \\ 
   & \emph{CT without delayed entry} & 0.620 & 0.972 & 0.000 &  & 0.898 \\ 
   & \emph{Ignoring external infection} & 0.620 & 0.972 & 0.000 &  &  \\ 
  \cmidrule{2-7}\multirow{4}{*}{Weibull} & Complete cohort & 0.935 & 0.936 & 0.077 & 0.072 & 0.918 \\ 
   & CT with delayed entry & 0.936 & 0.919 & 0.119 & 0.105 & 0.991 \\ 
   & \emph{CT without delayed entry} & 0.935 & 0.894 & 0.118 & 0.004 & 0.897 \\ 
   & \emph{Ignoring external infection} & 0.936 & 0.894 & 0.118 & 0.004 &  \\ 
   \bottomrule
\end{tabular}
\endgroup
\end{table}

\begin{table}[ht]
\centering
\caption{Proportion of fitted models with the lowest AIC} 
\label{tab:aic}
\begingroup\small
\begin{tabular}{llccc}
  \toprule
Contact intervals & Study design & Exponential & Weibull & Log-logistic \\ 
 \midrule
\multicolumn{5}{l}{\textbf{Who-infected-whom observed}}\\\multirow{2}{*}{Exponential} & Complete cohort & 0.727 & 0.105 & 0.168 \\ 
   & CT with delayed entry & 0.734 & 0.104 & 0.162 \\ 
  \cmidrule{2-5}\multirow{2}{*}{Log-logistic} & Complete cohort & 0.000 & 0.058 & 0.942 \\ 
   & CT with delayed entry & 0.000 & 0.059 & 0.942 \\ 
  \midrule
\multicolumn{5}{l}{\textbf{Who-infected-whom not observed}}\\\multirow{2}{*}{Exponential} & Complete cohort & 0.728 & 0.112 & 0.161 \\ 
   & CT with delayed entry & 0.734 & 0.110 & 0.157 \\ 
  \cmidrule{2-5}\multirow{2}{*}{Log-logistic} & Complete cohort & 0.000 & 0.059 & 0.942 \\ 
   & CT with delayed entry & 0.000 & 0.059 & 0.941 \\ 
   \bottomrule
\end{tabular}
\endgroup
\end{table}

\begin{table}[ht]
\centering
\caption{AIC values for regression models including all available covariates.} 
\label{tab:LAaic}
\begin{tabular}{l|ccc}
  \toprule
  \textbf{Internal} & \multicolumn{3}{c}{\textbf{External contact intervals}} \\
 \textbf{contact intervals} & Exponential & Weibull & Log-logistic \\
 \midrule
Exponential & 207.74 & 208.01 & 207.66 \\ 
  Weibull & 209.25 & 209.87 & 209.58 \\ 
  Log-logistic & 209.22 & 209.82 & 209.52 \\ 
   \bottomrule
\end{tabular}
\end{table}

\begin{table}[ht]
\centering
\caption{Coefficient estimates from final model with likelihood ratio confidence limits and p-values.} 
\label{tab:coefs}
\begin{tabular}{lrcr}
  \toprule
Coefficient & Estimate & 95\% confidence interval & p-value \\ 
  \midrule 
 \multicolumn{4}{l}{\textbf{Accounting for external infection}} \\
\texttt{intercept} & -4.90 & (-22.92, -3.38) & $< 0.001$ \\ 
  \texttt{adult\_inf} & 1.36 & (-0.36, 6.47) & 0.131 \\ 
  \texttt{adult\_sus} & -0.48 & (-1.46, 0.18) & 0.138 \\ 
  \texttt{proph\_sus} & -1.06 & (-2.26, -0.28) & 0.012 \\ 
  \texttt{xintercept} & -4.10 & (-6.04, -3.59) & 0.021 \\ 
  \texttt{xlogshape} & 0.80 & (-0.76, 1.50) & 0.191 \\ 
   [5pt] \multicolumn{4}{l}{\textbf{Ignoring external infection}} \\
\texttt{intercept} & -4.10 & (-5.46, -3.04) & $< 0.001$ \\ 
  \texttt{adult\_inf} & 0.76 & (-0.45, 2.10) & 0.218 \\ 
  \texttt{adult\_sus} & -0.77 & (-1.87, 0.37) & 0.178 \\ 
  \texttt{proph\_sus} & -0.83 & (-2.14, 0.30) & 0.155 \\ 
   \bottomrule
\end{tabular}
\end{table}

\begin{table}[ht]
\centering
\caption{Predicted household secondary attack risks with Wald confidence intervals.} 
\label{tab:sar}
\begin{tabular}{llrc}
  \toprule
  \textbf{Transmission} & & & \\
 \textbf{from} & \textbf{to} & Estimate & 95\% confidence interval \\
 \midrule 
 
 \multirow[c]{2}{*}{Child} & Child untreated & 4.4\% & (0.5\%, 34.6\%) \\ 
   & Child on prophylaxis & 1.5\% & (0.2\%, 13.7\%) \\ 
   [5pt] 
 \multirow[c]{2}{*}{Child} & Adult untreated & 2.7\% & (0.3\%, 20.1\%) \\ 
   & Adult on prophylaxis & 0.9\% & (0.1\%, 7.9\%) \\ 
   [5pt] 
 \multirow[c]{2}{*}{Adult} & Child untreated & 15.9\% & (6.4\%, 36.5\%) \\ 
   & Child on prophylaxis & 5.8\% & (2.0\%, 15.9\%) \\ 
   [5pt] 
 \multirow[c]{2}{*}{Adult} & Adult untreated & 10.2\% & (4.3\%, 22.8\%) \\ 
   & Adult on prophylaxis & 3.6\% & (1.2\%, 10.5\%) \\ 
   \bottomrule
\end{tabular}
\end{table}

\begin{table}[ht]
\centering
\caption{Log rate ratios with likelihood ratio confidence intervals and p-values from sensitivity analysis.} 
\label{tab:sensitivity}
\begin{tabular}{l|rcr|rcr}
  \toprule
  & \multicolumn{3}{c|}{\textbf{Accounting for external infection}} & \multicolumn{3}{c}{\textbf{Ignoring external infection}} \\
 Coefficient & Estimate & 95\% CI & p-value & Estimate & 95\% CI & p-value \\
 \midrule 
 \multicolumn{7}{l}{\textbf{Latent period = 1 day}} \\
\texttt{adult\_inf} & 0.08 & (-1.24, 1.37) & 0.894 & 0.06 & (-0.99, 1.09) & 0.909 \\ 
  \texttt{adult\_sus} & -0.63 & (-1.52, 0.07) & 0.074 & -0.58 & (-1.61, 0.52) & 0.286 \\ 
  \texttt{proph\_sus} & -1.05 & (-2.26, -0.23) & 0.013 & -0.95 & (-2.24, 0.13) & 0.087 \\ 
   [5pt] \multicolumn{7}{l}{\textbf{Infectious period = 5 days}} \\
\texttt{adult\_inf} & 1.73 & (-0.43, 4.96) & 0.140 & 0.58 & (-0.68, 1.95) & 0.366 \\ 
  \texttt{adult\_sus} & -0.41 & (-1.40, 0.20) & 0.167 & -0.91 & (-2.08, 0.25) & 0.121 \\ 
  \texttt{proph\_sus} & -1.02 & (-2.18, -0.29) & 0.011 & -0.69 & (-2.02, 0.47) & 0.246 \\ 
   [5pt] \multicolumn{7}{l}{\textbf{Infectious period = 7 days}} \\
\texttt{adult\_inf} & 0.29 & (-0.99, 1.73) & 0.649 & 0.20 & (-0.88, 1.28) & 0.707 \\ 
  \texttt{adult\_sus} & -0.57 & (-1.38, 0.15) & 0.103 & -0.44 & (-1.44, 0.64) & 0.406 \\ 
  \texttt{proph\_sus} & -0.90 & (-1.96, -0.12) & 0.026 & -0.73 & (-1.89, 0.30) & 0.167 \\ 
   \bottomrule
\end{tabular}
\end{table}

\end{document}